\pdfoutput=1
\documentclass[]{JFM-FLM_Au} 

\usepackage{multirow}

\lefttitle{Ding Wang et al.}
\righttitle{Journal of Fluid Mechanics}

\title{Symbolic Regression-Enhanced Dynamic Wake Meandering: Fast and Physically Consistent Wind-Turbine Wake Modeling}

\author{Ding Wang\aff{1,2}, Dachuan Feng\aff{4}, Kangcheng Zhou\aff{5}, Yuntian Chen\aff{2,3}, Shijun Liao\aff{1} \and Shiyi Chen\aff{2,3}}

\affiliation{\aff{1}School of Ocean and Civil Engineering, Shanghai Jiao Tong University, Shanghai, China
\aff{2}Ningbo Key Laboratory of Advanced Manufacturing Simulation, Eastern Institute of Technology, Ningbo, China
\aff{3}Ningbo Institute of Digital Twin, Eastern Institute of Technology, Ningbo, China
\aff{4}Faculty of Aerospace Engineering, Delft University of Technology, Delft, Netherlands
\aff{5}Department of Mechanical Engineering, The University of Hong Kong, Hong Kong, China}

\begin{document}
\maketitle

\begin{abstract}
Accurately modeling wind turbine wakes is essential for optimizing wind farm performance but remains a persistent challenge. While the dynamic wake meandering (DWM) model captures unsteady wake behavior, it suffers from near-wake inaccuracies due to empirical closures. We propose a Symbolic Regression-enhanced DWM (SRDWM) framework that achieves equation-level closure by embedding symbolic expressions for volumetric forcing and boundary terms explicitly into governing equations. These physically consistent expressions are discovered from LES data using symbolic regression guided by a hierarchical, domain-informed decomposition strategy. A revised wake-added turbulence formulation is further introduced to enhance turbulence intensity predictions. Extensive validation across varying inflows shows that SRDWM accurately reproduces both mean wake characteristics and turbulent dynamics, achieving full spatiotemporal resolution with over three orders of magnitude speedup compared to LES. The results highlight symbolic regression as a bridge between data and physics, enabling interpretable and generalizable modeling.
\end{abstract}



\section{Introduction}\label{sec:intro}
Wind turbine wakes are central to power loss and flow interactions across wind farms. Yet accurately and efficiently capturing their unsteady behavior remains a major challenge in wind energy research. Most engineering wake models focus on steady-state predictions for long-term power estimation~\citep{WD_hill}, overlooking key unsteady features such as meandering, intermittency, and coherent structures. These dynamics significantly affect power fluctuations and structural loads, and are critical for instantaneous power estimation, dynamic control design, and grid frequency regulation~\citep{milan2013turbulent,stevens2017flow}. Efficient models capable of resolving the full spatiotemporal evolution of wakes are thus vital for next-generation wind farm design and operation. 

In this context, the dynamic wake meandering (DWM) model was proposed~\citep{larsen2008wake}, offering a framework to incorporate unsteady effects in engineering-focused simulations. This model treats the wake as a passive tracer whose large-scale motion is driven by atmospheric boundary layer (ABL) turbulence, thus capturing the temporal displacement of the wake. The physical foundation of DWM traces back to the early hypothesis by~\cite{lissaman1983wake}, which identified large-scale atmospheric eddies as the dominant driver of wake meandering—a view supported by wind tunnel experiments and field measurements~\citep{espana2011spatial,trujillo2011light}. Over time, the DWM model has been extended to incorporate more physical effects, including ABL shear~\citep{keck2015two}, yaw misalignment~\citep{branlard2023time}, and wake-added turbulence (WAT) modeling~\citep{branlard2024development}. Model parameters have been systematically calibrated using experimental and numerical approaches~\citep{madsen2010calibration,doubrawa2018optimization}, and different DWM variants have been compared and evaluated~\citep{hanssen2023comparison}. These mid-fidelity models have demonstrated reliable accuracy in far-wake predictions when validated against large-eddy simulations (LES) results~\citep{shaler2021fast,rivera2024comparison} and field data~\citep{kretschmer2021fast,shaler2024wind}, making them suitable for layout design and power forecasting.

Despite these advances, the physical origin of wake meandering remains debated. While the classic DWM formulation attributes wake displacement to large-scale turbulence, an alternative perspective highlights the role of intrinsic shear-layer instabilities, as evidenced by laboratory experiments~\citep{medici2006measurements,okulov2014regular} and numerical studies~\citep{kang2014onset}. Recent work suggests a coexistence of both mechanisms~\citep{heisel2018spectral,yang2019wake}, where large-scale turbulence amplifies inherent wake instabilities~\citep{coudou2017experimental}. Building on this dual-mechanism hypothesis, subsequent extensions to the DWM framework have incorporated both effects by selectively amplifying turbulence fluctuations, leading to improved predictions of wake-added turbulence intensity~\citep{DC_FF}.

Nevertheless, most existing DWM models still lack accurate treatment of near-wake dynamics. Like traditional steady-state models, they often prioritize far-wake accuracy at the expense of physical consistency in the near wake. Major deficiencies include the omission of pressure recovery and the artificial prescription of wake expansion onset, leading to physically incomplete calculations. To improve near-wake performance, steady-state models have adopted self-similar double-Gaussian (DG) velocity deficit profiling~\citep{keane2016analytical}, with extensions that enhance mass conservation~\citep{schreiber2020brief,keane2021advancement} and validate performance against LES~\citep{soesanto2022anisotropic}. More recently, data-driven variants of DG models~\citep{WD_SR} have achieved improved agreement with high-fidelity data through refined profile fitting. However, these approaches remain fundamentally limited: they treat near-wake modeling as a curve-fitting task, without addressing the deeper issue of how the underlying governing equations should be closed. Consequently, while they may reproduce mean velocity profiles, they lack dynamical consistency—particularly under unsteady or complex inflow conditions. More importantly, the fitted profiles offer limited physical insight, whereas achieving equation-level closure can potentially reveal interpretable mechanisms and inform the development of physically grounded wake models.

Symbolic regression (SR) offers a promising solution to this modeling gap. As an interpretable machine learning technique, SR can automatically discover concise and accurate mathematical expressions that describe complex nonlinear relationships in data~\citep{angelis2023artificial}. Compared to black-box data-driven methods like neural networks, SR is better suited for explicitly uncovering governing physical laws~\citep{makke2024interpretable}. Mainstream approaches include sparse identification of nonlinear dynamics (SINDy)~\citep{brunton2016discovering} and genetic programming with expression trees~\citep{chen2022symbolic}, which have shown strong performance across applications~\citep{la2021contemporary}. Recent advances in SR algorithms~\citep{kamienny2022end,tenachi2023deep,du2024discover} has further accelerated its adoption in data-driven physical modeling~\citep{gamella2025causal}. In fluid dynamics, SR has recently demonstrated success in modeling flow behavior with both accuracy and interpretability~\citep{ma2024dimensional,WD_SR}.

To address the limitations of current models, we propose a Symbolic Regression-enhanced Dynamic Wake Meandering (SRDWM) framework. The first contribution is the integration of symbolic expressions, extracted from LES data, into the DWM governing equations to reconstruct volumetric forcing and boundary terms—achieving equation-level closure and restoring near-wake dynamics. Second, we introduce a hierarchical SR strategy guided by domain knowledge to reduce dimensionality and improve generalizability. Additionally, SRDWM incorporates a revised WAT formulation to enhance wake turbulence predictions under the dual-mechanism hypothesis of wake meandering. Together, these developments yield a model recovers physical consistency in both space and time, capturing unsteady wake evolution with high accuracy and computational efficiency.

The rest of the paper is structured as follows. Section~\ref{sec:methods} details the inflow generation, wake simulation, and the proposed SRDWM framework. Section~\ref{sec:results} presents the discovered symbolic terms and evaluates wake dynamics across inflow conditions. Finally, we draw a conclusion in section~\ref{sec:conclusion}.

\section{Methods}\label{sec:methods}
\begin{figure*}
    \centering
    \includegraphics[width=\linewidth]{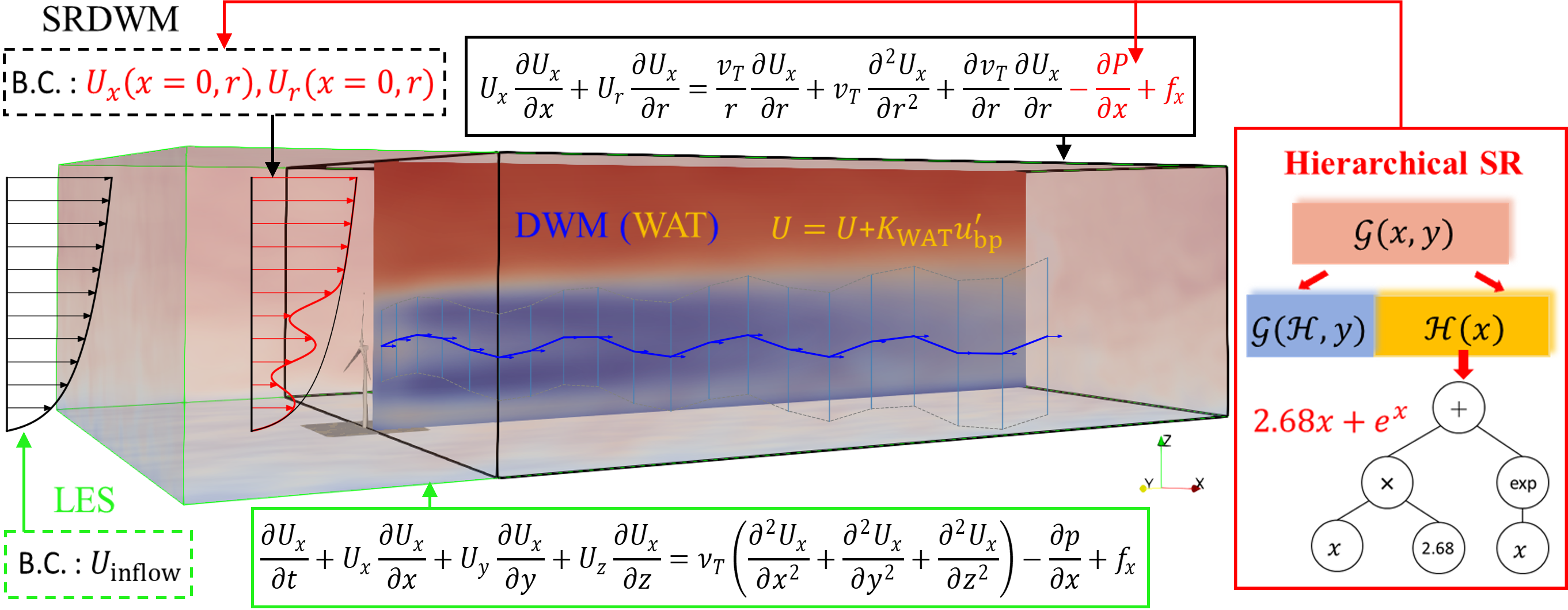}
    \caption{Schematic of the SRDWM modeling framework. The proposed method (black) augments the DWM model (blue) with hierarchical SR (red) to reconstruct volumetric forcing and wake boundary conditions from data of LES (green), and improves predictions of wake dynamics with the WAT model (yellow).}
    \label{fig:schematic}
\end{figure*}

This section introduces the SRDWM framework (see figure~\ref{fig:schematic}), which enhances DWM by embedding symbolic expressions—reconstructed from LES data via hierarchical symbolic regression—into the governing equations. These replace empirical terms for forcing and boundary conditions, improving physical consistency. A revised WAT model further refines wake turbulence predictions. Together, these components yield an accurate, efficient, and interpretable wake model with full spatiotemporal resolution.

\subsection{Inflow turbulence generation}\label{subsec:inflow}
For LES and DWM-type models, turbulence characteristics in the incoming flow strongly influence wake recovery and meandering behavior~\citep{yang2019review}. In this study, the inflow turbulence, serving as the inlet boundary condition for turbine wakes, is generated using two approaches: precursor simulation (PS) and synthetic turbulence (ST). Both methods produce ABL turbulence with identical logarithmic mean streamwise velocity profiles, but differ in turbulence intensity. Specifically, the PS case exhibits a hub-height turbulence intensity (TI) of 6.1\%, while the ST cases span a broader range: 1.8\% (ST1), 4.1\% (ST2), 6.4\% (ST3), 8.4\% (ST4), and 10.3\% (ST5).


In the PS approach, a fully developed turbulent ABL flow is simulated using LES with periodic horizontal boundary conditions. Once statistical steady state is reached, the velocity fields at the inlet are recorded at each time step for use in subsequent turbine-wake simulations.

Alternatively, the ST method generates inflow turbulence by modeling a stationary stochastic process governed by prescribed cross-spectral density. Initially, the Kaimal spectrum~\citep{kaimal} is applied to define the turbulence energy distribution in the frequency domain, ensuring that the resulting velocity time series adheres to the target power spectral density (PSD). The turbulence spectrum, following the IEC 61400-1 standard, is expressed as 
\begin{equation}
    \frac{f S_{\chi}(f)}{\sigma_{\chi}^2} = \frac{4 f L_{\chi} / \bar{U}_{\text{hub}}}{(1 + 6 f L_{\chi} / \bar{U}_{\text{hub}})^{5/3}},
\end{equation}  
where $S_{\chi}$ denotes the auto-spectral density function, $\chi \in \{x,y,z\}$ represents directions, $\sigma_{\chi}$ is the standard deviation of velocity components, $L_{\chi}$ is the integral length scale, $f$ is the cyclic frequency, and $\bar{U}_{\text{hub}}$ is the mean hub-height velocity. To further account for spatial coherence, a model is incorporated to characterize the correlation properties of turbulence fluctuations:  
\begin{equation}
    \Theta(r, f) = \exp \left( -12 \sqrt{\left(\frac{f r}{\bar{U}_{\text{hub}}}\right)^2 + \left(\frac{0.12 r}{L_c}\right)^2 } \right),
\end{equation}  
where $\Theta(r,f)$ quantifies coherence between two nodes separated by distance $r$ (perpendicular to wind direction) and $L_c$ is the coherence length scale. Synthesized turbulence fluctuations are subsequently superimposed onto the mean velocity profile, forming spatiotemporally correlated wind fields. The resulting turbulence data undergoes inverse Fourier transformation and is structured as a time series along a two-dimensional boundary, establishing the inflow turbulence conditions.  

\subsection{Turbine wake simulation}\label{subsec:turWakeSim}
Utilizing the numerical solver "SOWFA" \citep{SOWFA}, LES is performed to simulate wind turbine wakes under varying inflow turbulence intensities. The turbine effect is represented using the actuator disk method with rotation (ADM-R), a widely validated approach in previous studies~\citep{porte2011large,stevens2018comparison}. The wake dynamics are governed by the filtered incompressible Navier–Stokes (NS) equations:
\begin{equation}
\frac{\partial{\widetilde{U}_j}}{\partial{x_j}}=0,
\end{equation}
\begin{equation}
\frac{\partial \widetilde{U}_i}{\partial t}+\frac{\partial(\widetilde{U}_i\widetilde{U}_j)}{\partial{x_j}}=-\frac{\partial \widetilde{p}^*}{\partial x_i}-\frac{\partial \tau_{i j}^D}{\partial x_j}+\frac{1}{\rho} F_i^T.
\label{eq:momen}
\end{equation}
Here, $\widetilde{U}_{i}$ denotes the filtered velocity component, where $i = 1, 2, 3$ corresponds to the streamwise ($x$), spanwise ($y$), and vertical ($z$) directions, respectively. The term $\widetilde{p^*}$ represents the modified filtered kinematic pressure, and $\tau_{ij}^D$ is the deviatoric component of the sub-grid (SGS) scale stress. The SGS effects are parameterized using the Lagrangian-averaged scale-dependent dynamic model~\citep{2005LASD}, which dynamically computes anisotropic eddy viscosity with scale-adaptive Smagorinsky coefficients. The turbine force term $F_i^T$ is derived by spatially smoothing the aerodynamic forces exerted by the turbine blades via a convolution operation:
\begin{equation}
F_i^T = \frac{\boldsymbol{f}}{\epsilon^3 \pi^{3/2} dV} \exp \left(-\left(\frac{d}{\epsilon}\right)^2  \right),
\end{equation}
where $dV$ is the volume element, $\epsilon = 20$~m controls the smoothing effect, and $d$ represents the distance between the actuator point and the computational grid center. This regularization ensures numerical stability and a smoothly distributed turbine force.  

The aerodynamic force $\boldsymbol{f}$, acting on the blade elements, is computed using blade element momentum theory:
\begin{equation}
\boldsymbol{f}=\frac{1}{2} \rho U_{\text{rel}}^2c dr (C_L \boldsymbol e_L+C_D \boldsymbol e_D),
\label{eq:ADM}
\end{equation}  
where ${U}_{\text{rel}}$ represents the relative velocity, $c$ is the blade-chord length, and $dr$ is the radial width. The lift ($C_L$) and drag ($C_D$) coefficients are interpolated from aerodynamic lookup tables, while $\boldsymbol e_L$ and $\boldsymbol e_D$ denote unit vectors.  

The wind turbine modeled in this study is the NREL 5MW reference turbine, characterized by a $126$~m rotor diameter (D) and $90$~m hub height. It operates at $9$ rotations per minute, maintaining an optimal tip-speed ratio. Following prior studies~\citep{liu2020effects,yang2022effects}, the computational domain, spanning $1800\times1000\times600$~m$^3$, is meshed using a structured grid comprising $180\times100\times80$ nodes in the $x$-, $y$-, and $z$-directions. Periodic boundary conditions are enforced in the spanwise directions, while a wall model resolves near-ground dynamics. Velocity at the fourth grid point is adjusted to correspond to the wall shear stress, reducing discrepancies in the logarithmic layer~\citep{2012Wall}. The accuracy of the simulated flow fields has been validated against high-fidelity data~\citep{chen2025three}, with further computational setup and solver implementation detailed in Refs.~[\citenum{DC_component,WD_hill}].

\subsection{Symbolic Regression-enhanced Dynamic Wake Meandering model}\label{subsec:SRDWM}

\subsubsection{DWM framework}\label{subsubsec:DWM}
The DWM model hypothesizes that wake behavior resembles that of a passive tracer advected by large-scale atmospheric turbulence~\citep{larsen2008wake}. Under this assumption, the model applies a thin shear layer approximation of the NS equations~\citep{ainslie1988calculating} to compute the velocity deficit in a meandering frame, where the deficit is assumed quasi-steady relative to the instantaneous wake center. Wake meandering is then introduced by advecting the deficit pattern along the pathlines determined by the low-pass filtered velocity field.

\begin{figure*}
    \centering
    \includegraphics[width=\linewidth]{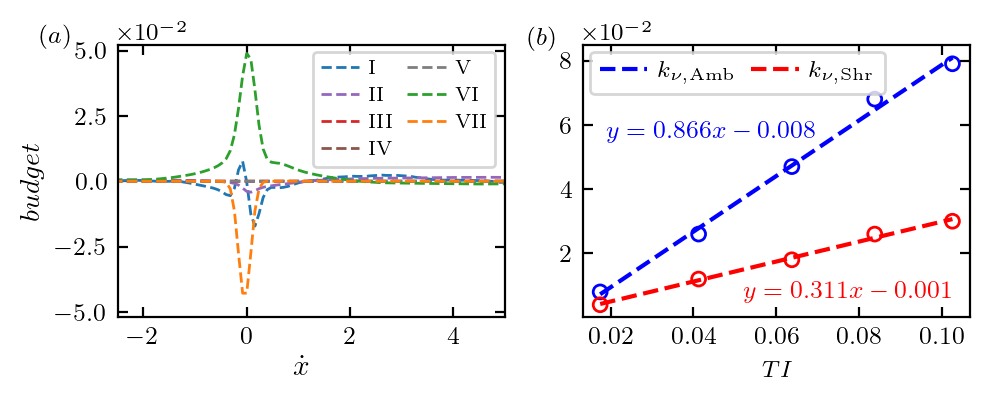}
    \caption{(a) Budget of the mean momentum equation normalized by $U_\infty^2$ for the LES PS case. (b)Linear fitting of coefficients $k_{\nu, \mathrm{Amb}}$ and $k_{\nu, \mathrm{Shr}}$ versus turbulence intensity $TI$.}
    \label{fig:paramK}
\end{figure*}

The original governing equations neglect pressure gradient effects and assumes dominance of radial over axial velocity gradients in the axisymmetric formulation. However, as shown in figure~\ref{fig:paramK}(a), LES results reveal that both mean pressure gradient (term VI) and mean turbine-induced forcing (term VII) make significant contributions, particularly in the near-wake region. We therefore restore these terms, leading to the following momentum and continuity equations:
\begin{equation}
\underbrace{r U_x \frac{\partial U_x}{\partial x}}_{\substack{\text{I} \\ \text{Axial advection}}} + 
\underbrace{r U_r \frac{\partial U_x}{\partial r}}_{\substack{\text{II} \\ \text{Radial advection}}} = \underbrace{\nu_T \frac{\partial U_x}{\partial r} + r \nu_T \frac{\partial^2 U_x}{\partial r^2} + r \frac{\partial \nu_T}{\partial r} \frac{\partial U_x}{\partial r}}_{\substack{\text{III–V} \\ \text{Radial diffusion}}}- \underbrace{r \frac{\partial P}{\partial x}}_{\substack{\text{VI} \\ \text{Pressure gradient}}} + 
\underbrace{r f_x}_{\substack{\text{VII} \\ \text{Turbine forcing}}},
\label{eq:TSLA}
\end{equation}
\begin{equation}
U_r + r \frac{\partial U_r}{\partial r} + r \frac{\partial U_x}{\partial x} = 0.
\end{equation}
Here, $U_x$ and $U_r$ denote the axial and radial velocity components, $r$ stands for the radial position, and $\nu_T$ represents the eddy viscosity. Following~\cite{doubrawa2018optimization}, $\nu_T$ is modeled as the sum of ambient velocity and wake shear contributions:
\begin{equation}
\begin{aligned}
\nu_T(x, r) =\;& \mathcal{F}_{\nu, \mathrm{Amb}}(x)\, k_{\nu, \mathrm{Amb}}\, \widehat{TI}_{\text{Amb}}\, R\, \widehat{U}_{x,\text{Amb}} \\
&+ \mathcal{F}_{\nu, \mathrm{Shr}}(x)\, k_{\nu, \mathrm{Shr}}\, \max\left( R^2 \left| \frac{\partial U_x}{\partial r} \right|, R \min_r U_x \right),
\end{aligned}
\end{equation}
where $\widehat{\cdot}$ denotes low-pass temporal filtering, $R$ is the rotor radius, and the filter functions $\mathcal{F}$ account for the initial imbalance between velocity and turbulence fields, with parameters calibrated in prior studies~\citep{madsen2010calibration,keck2015two,doubrawa2018optimization}. The coefficients $k_{\nu,\mathrm{Amb}}$ and $k_{\nu,\mathrm{Shr}}$ associated with dissipation are linearly regressed from the turbulence intensity $TI$, as shown in figure~\ref{fig:paramK}(b).

To improve the underestimation of turbulence intensity in the original DWM formulation, the WAT model~\citep{madsen2010calibration} is commonly employed. Building upon this, \citet{DC_FF} proposed a scale-dependent amplification mechanism based on shear instability, realizing the dual-mechanism hypothesis. While this WAT model enhances spectral predictions, it also indiscriminately amplifies velocity fluctuations across the entire wake. As a result, it leads to overpredictions of near-wake turbulence intensity, which has been shown to exhibit localized suppression~\citep{heisel2018spectral,li2024impacts}.

To resolve this issue, we propose a revised amplification scheme using a wake-scaling factor $K_{\text{WAT}}(x, r)$ that selectively enhances band-pass filtered fluctuations. These filtered components, denoted by $\boldsymbol{u}^{\prime}_{\mathrm{bp}}$, are extracted through recursive exponential smoothing, combining low-pass $\boldsymbol{u}_{\mathrm{lp}}$ and high-pass $\boldsymbol{u}^{\prime}_{\mathrm{hp}}$ signals as follows:
\begin{equation}
\begin{aligned}
\boldsymbol{u}^{\prime}_{\mathrm{bp}}[n+1] &= \beta\, \boldsymbol{u}^{\prime}_{\mathrm{bp}}[n] + (1 - \beta)\, \boldsymbol{u}^{\prime}_{\mathrm{hp}}[n+1], \\
\boldsymbol{u}^{\prime}_{\mathrm{hp}}[n+1] &= \boldsymbol{u}[n] - \boldsymbol{u}_{\mathrm{lp}}[n], \\
\boldsymbol{u}_{\mathrm{lp}}[n+1] &= \alpha\, \boldsymbol{u}_{\mathrm{lp}}[n] + (1 - \alpha)\, \boldsymbol{u}[n+1],
\end{aligned}
\end{equation}
where $\boldsymbol{u}$ denotes instantaneous velocity and $\boldsymbol{u}^{\prime}$ its fluctuation. The filtering coefficients are defined as $\alpha = e^{-2\pi \Delta t f_1}$ and $\beta = e^{-2\pi \Delta t f_2}$, with cutoff frequencies $f_1$ and $f_2$ spanning the energy-dominant Strouhal ($St=fD/U_{\infty}$) band $0.1 < St < 1.0$~\citep{DC_component}.

The refined wake-scaling factor is defined as:
\begin{equation}
K_{\mathrm{WAT}}(x, r) = \frac{\mathcal{F}_{s,\mathrm{Amb}}(x) \, k_{s,\mathrm{Amb}}}{U_{x,\mathrm{Amb}}} \, \frac{\partial U_x}{\partial r} - \mathcal{F}_{s,\mathrm{Shr}}(x) \, k_{s,\mathrm{Shr}} \left| 1 - \frac{U_x}{U_{x,\mathrm{Amb}}} \right|,
\label{eq:K}
\end{equation}
with $k_{s,\mathrm{Amb}}=3.5$ and $k_{s,\mathrm{Shr}}=2.3$. Unlike previous formulations, this structure—accounting for negative velocity gradients near the centerline—enhances shear-driven scaling while suppressing unphysical growth in the near wake, where the velocity deficit peaks. By amplifying coherent band-pass fluctuations, this approach yields more accurate turbulence intensity distributions, as will be demonstrated in section~\ref{sec:results}.

\subsubsection{Symbolic regression}\label{subsubsec:SR}
\begin{figure}
    \centering
    \includegraphics[width=\linewidth]{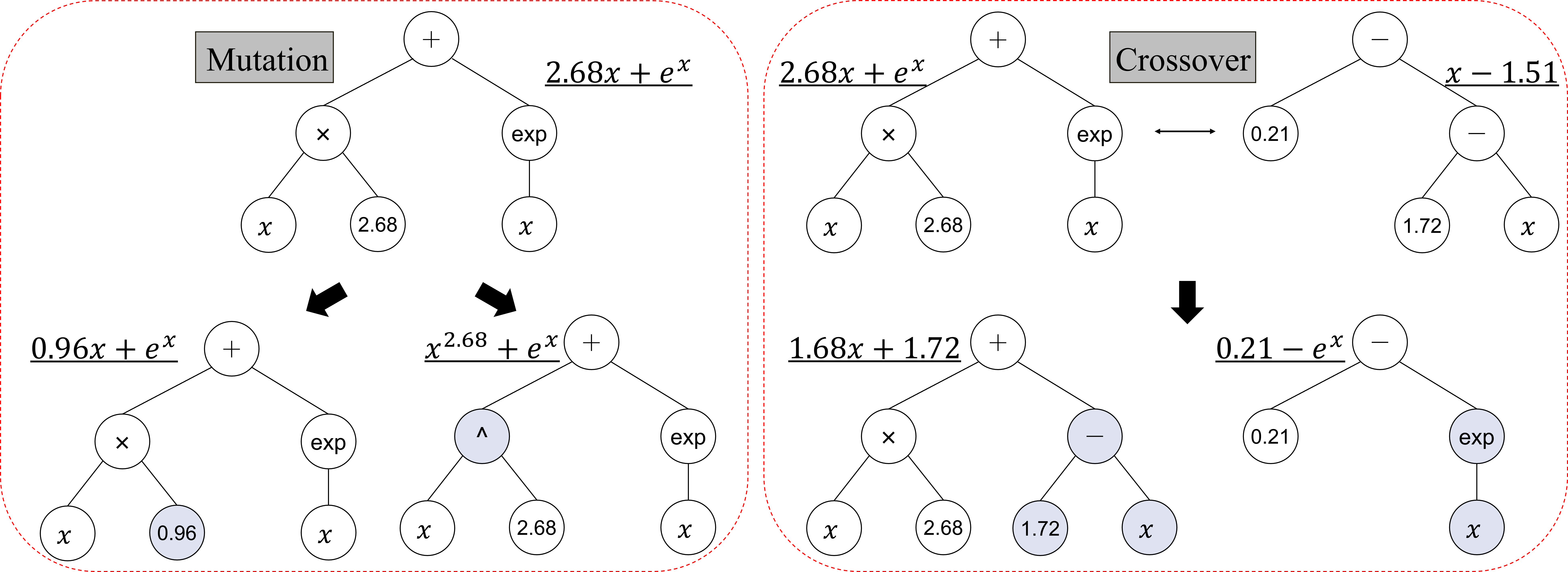}
    \caption{Schematic of evolution operations in the genetic algorithm-based SR.}
    \label{fig:SR_schematic}
\end{figure}

SR is a data-driven approach for discovering interpretable equations that explicitly capture complex nonlinear relationships between variables, providing concise mathematical descriptions of physical system behavior. It systematically explores combinations of variables, constants, and mathematical operators, while balancing descriptive accuracy with model simplicity. In this study, we adopt the PySR library~\citep{cranmer2023interpretable}, which evolves expression trees—composed of operators (nodes) and variables or constants (leaves)—using a multi-population genetic algorithm. The evolutionary process includes selection, crossover, and mutation, as illustrated schematically in figure~\ref{fig:SR_schematic}. Selection is governed by asynchronous tournament strategies~\citep{goldberg1991comparative}, while inter-population migration enhances diversity and prevents premature convergence. Simulated annealing~\citep{kirkpatrick1983optimization} is integrated to improve exploration, and the final symbolic expressions are refined through BFGS-based constant optimization~\citep{broyden1970convergence}.

\begin{figure*}
    \centering
    \includegraphics[width=\linewidth]{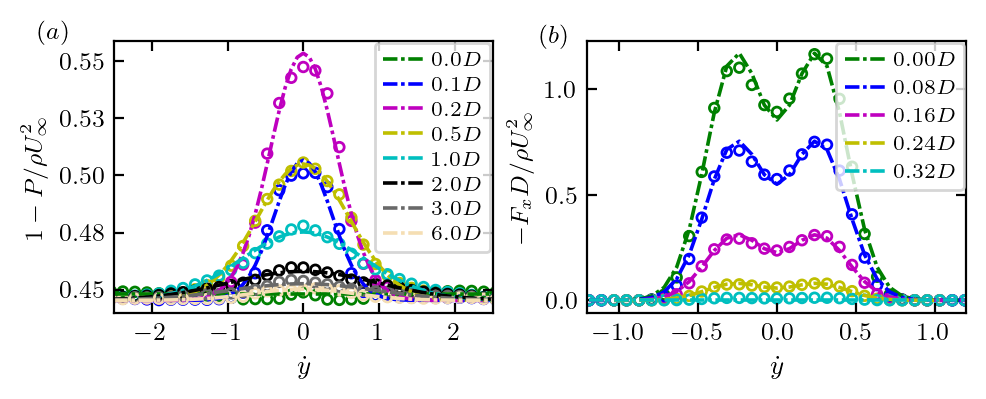}
    \caption{Spanwise distribution of (a) mean pressure and (b) mean turbine force at various downstream positions in the PS case. Circles show LES results, while dashdotted lines denote fitted Gaussian/DG profiles.}
    \label{fig:BF_similarity}
\end{figure*}

Given the limited availability of CFD data in wake simulations, we adopt a hierarchical regression strategy guided by domain knowledge. Specifically, we assume that pressure and turbine forcing terms exhibit separable forms: the spanwise distribution is prescribed a priori using physically motivated functions (equations~\ref{eq:P} and \ref{eq:Fx}), while the streamwise evolution is inferred through SR. As shown in figure~\ref{fig:BF_similarity}, these two forcing terms exhibit self-similar behavior, well captured by Gaussian and DG profiles, respectively: 
\begin{equation} 1-\frac{P}{\rho U_{\infty}^2} = a \times \exp\left(-\frac{\dot{y}^2}{2\sigma^2}\right), 
\label{eq:P}
\end{equation} 
\begin{equation} -\frac{F_x D}{\rho U_{\infty}^2} = a \times \left(\exp\left(-\frac{(\dot{y} - \mu)^2}{2\sigma^2}\right) + \exp\left(-\frac{(\dot{y} + \mu)^2}{2\sigma^2}\right)\right), 
\label{eq:Fx}
\end{equation} 
where $\dot{x} = x/D$ and $\dot{y} = y/D$ denote normalized coordinates. The parameters $a(\dot{x})$, $\mu(\dot{x})$, and $\sigma(\dot{x})$ describe the streamwise variation and are treated as independent scalar functions to be identified.
This formulation transforms the original two-dimensional regression task into one-dimensional sub-problems, forming a hierarchical model structure. By decoupling the spatial dimensions based on physical insight, the SR process becomes more tractable with significantly reduced searching space and the resulting expressions more interpretable.

To perform SR, we construct a candidate operator library consisting of basic arithmetic operations, powers, exponentials, and cosine functions—operators commonly encountered in wake modeling. To ensure parsimony and prevent overfitting, structural constraints are imposed on operator nesting (e.g., disallowing nested exponentials). Each of these streamwise parameters ($a$, $\mu$, $\sigma$) is regressed independently using the same input variable $\dot{x}$, with 20,000 iterations to yield concise symbolic expressions.

All expressions are derived exclusively from the PS case, which serves as the training dataset. The remaining ST cases are used for validation to assess the generalizability of the discovered expressions. These expressions explicitly encode the streamwise evolution of the volumetric forcing, and are directly embedded into the DWM numerical solver, forming the physical core of the SRDWM framework.

\section{Results and discussion}\label{sec:results}
This section presents a comprehensive evaluation of SRDWM performance against LES, treated as ground truth (GT), under various inflow conditions. Section~\ref{subsec:terms} presents the extracted symbolic expressions, section~\ref{subsec:wake} analyzes key features of wake evolution, and section~\ref{subsec:turbulence} further examines turbulence statistics and wake dynamics. Results show that SRDWM offers high accuracy and generalizability with full spatiotemporal resolution.

\subsection{SR-discovered volumetric terms and boundary conditions}\label{subsec:terms}
\begin{figure*}
    \centering
    \includegraphics[width=\linewidth]{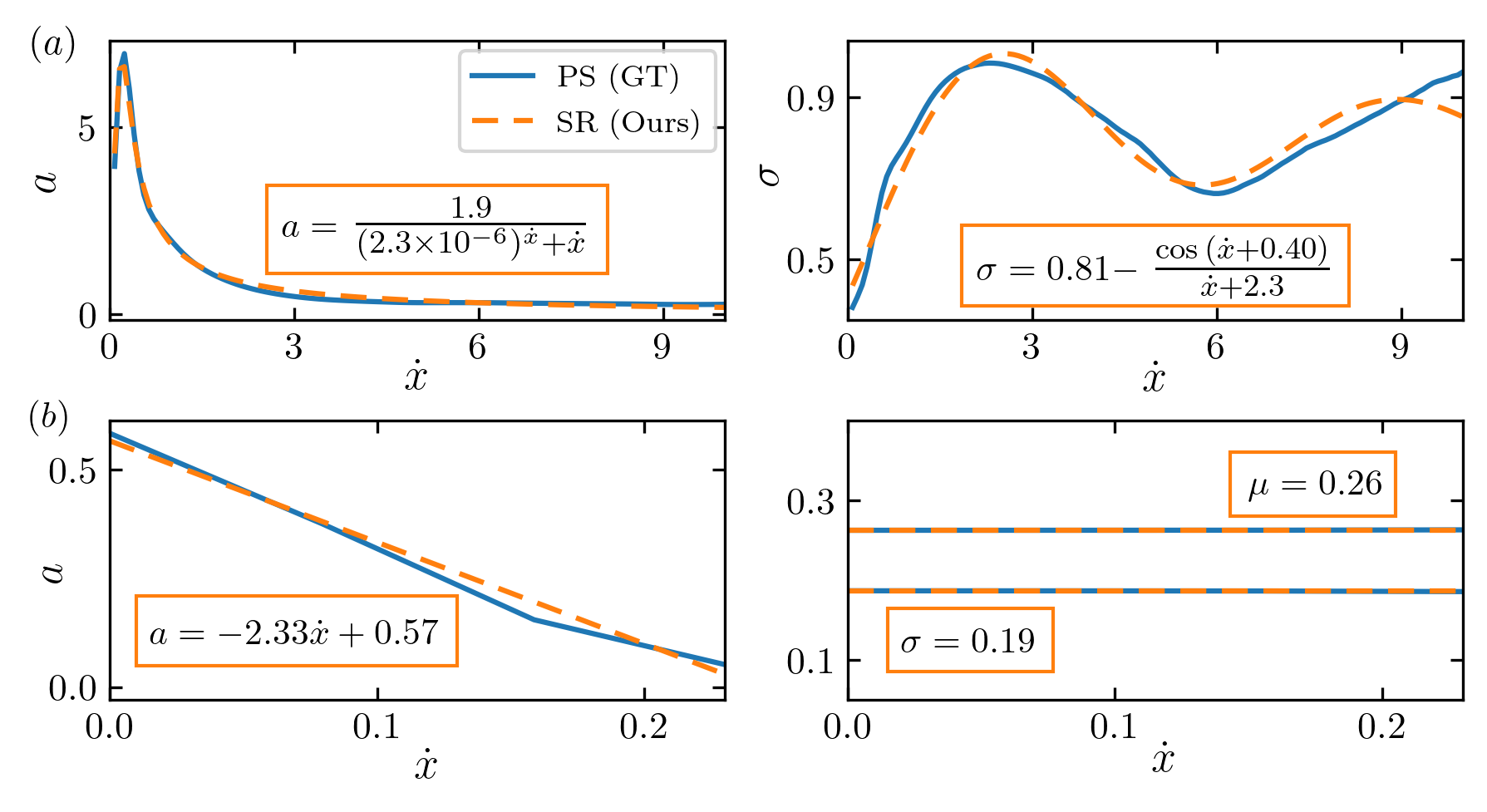}
    \caption{SR-discovered expressions for streamwise variation of (a) pressure and (b) turbine force terms.}
    \label{fig:expressions}
\end{figure*}

Figure~\ref{fig:expressions} displays the symbolic expressions discovered from the PS case for parameters $a$, $\mu$, and $\sigma$ of volumetric terms in equations~\ref{eq:P} and \ref{eq:Fx}. Remarkably, the amplitude function of the pressure term includes an exponential decay factor, capturing the localized pressure drop region behind the rotor. This indicates that pressure recovery occurs downstream of the actuator region rather than immediately behind the turbine, in agreement with physical expectations. For the turbine force distribution, the extracted DG profile shows a near-linear streamwise decay of amplitude, with constant mean and standard deviation. This behavior reflects how momentum loss from turbine extraction gradually weakens downstream while maintaining a consistent spanwise pattern. These findings validate the accuracy and physical interpretability of the SR-derived expressions.

\begin{figure*}
    \centering
    \includegraphics[width=1\linewidth]{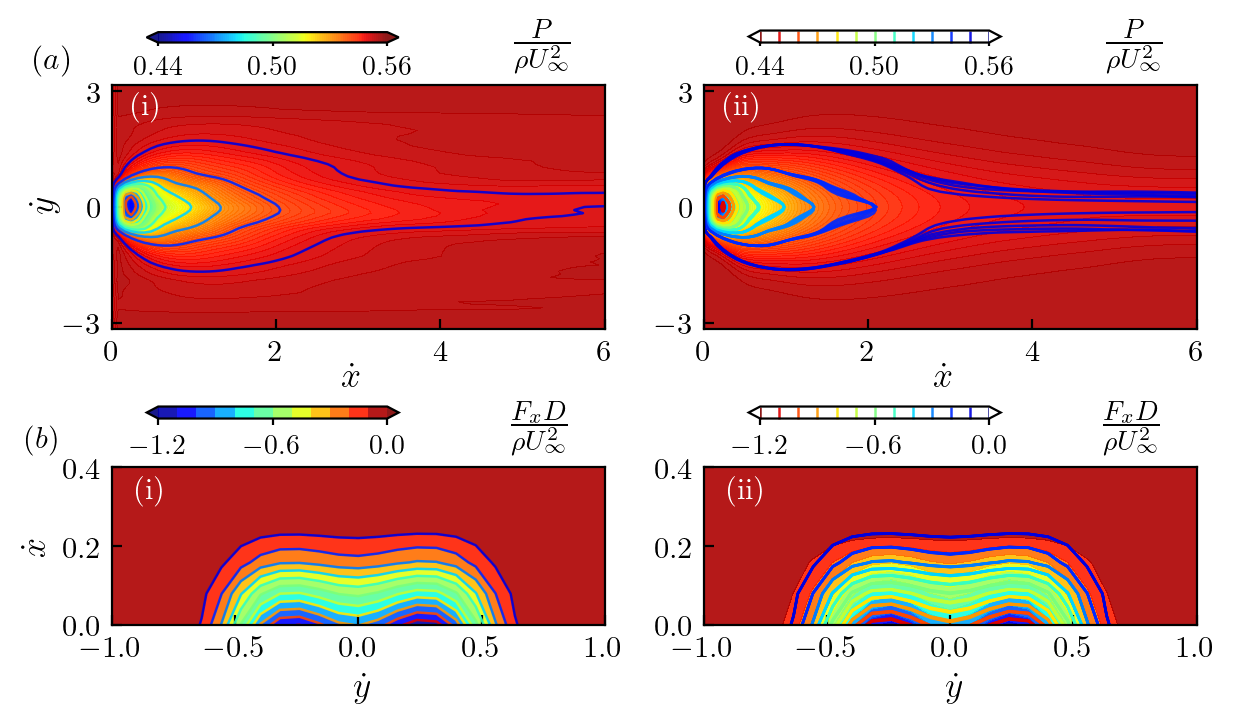}
    \caption{Distribution of (a) mean pressure and (b) mean turbine force: (i) LES colormaps and isocontours for PS case; (ii) SR predictions for PS case (colormap) with isocontours from all LES ST cases.}
    \label{fig:bodyForce}
\end{figure*}

Figure~\ref{fig:bodyForce} further assesses generalization performance by comparing the reconstructed pressure and turbine forcing in both the PS case (training data) and ST cases (test data). In figure~\ref{fig:bodyForce}(i), background colormaps show LES data for the PS case, while figure~\ref{fig:bodyForce}(ii) overlays the SR predictions with LES isocontours for all ST cases. The close alignment across all cases demonstrates that near-wake pressure and turbine forcing are governed primarily by turbine aerodynamics and remain largely invariant with respect to inflow turbulence. Far-wake variations, while present, are minor relative to the dominant near-wake features. This result highlights the robustness and strong generalization capacity of the SR-discovered equations for volumetric foring.

\begin{figure*}
    \centering
    \includegraphics[width=1\linewidth]{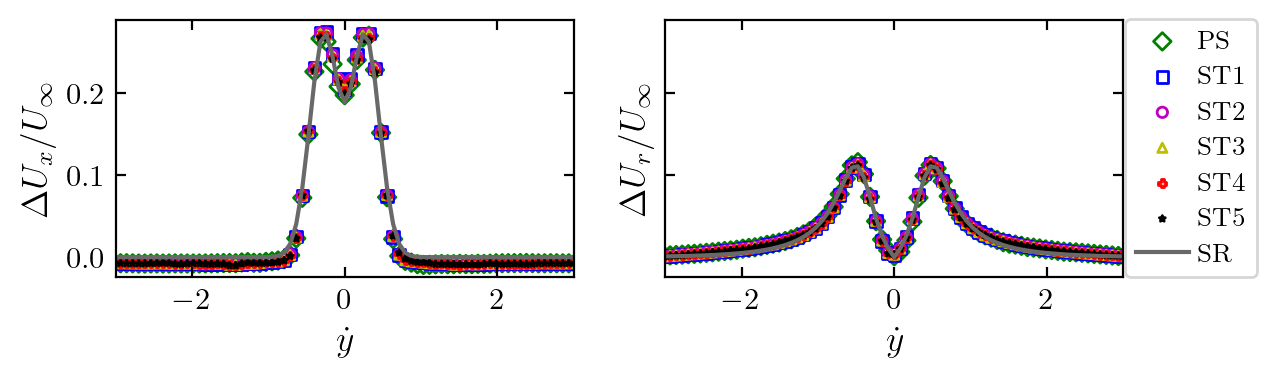}
    \caption{Comparison of mean wake velocity boundary conditions derived from SR and LES under different inflow turbulence intensities.}
    \label{fig:BC}
\end{figure*}

For the wake velocity boundary conditions used in SRDWM, we similarly decompose the velocity profile into fixed spanwise functions, with streamwise position set to the rotor location ($\dot{x} = 0$). The streamwise and radial velocity components are discovered respectively as:
\begin{equation}
\frac{\Delta U_x(\dot{x}=0)}{U_{\infty}} = 2.15 \times \left(\exp\left(-\frac{(\dot{r} -0.27)^2}{2\times 0.19^2}\right)+\exp\left(-\frac{(\dot{r}+0.27)^2}{2\times 0.19^2}\right) \right), 
\label{eq:U}
\end{equation}
\begin{equation}
\frac{\Delta U_r(\dot{x}=0)}{U_{\infty}} = \frac{0.88\dot{r}}{e^{2.0\dot{r}} - 4.4\dot{r}}.
\label{eq:V}
\end{equation}
As shown in figure~\ref{fig:BC}, the SR-derived expressions accurately reconstruct the initial mean wake velocity distribution across varying inflow turbulence intensities. The spanwise location for peak of streamwise velocity deficit aligns with that of the turbine forcing, and both variables remain largely insensitive to inflow variations. This consistency reflects the underlying blade aerodynamics~\citep{keane2021advancement}, particularly the lift distribution peaking near mid-span~\citep{Lift}. These forces are projected onto the mesh via Gaussian smoothing~\citep{SOWFA}, and iteratively generate the observed velocity profile through the momentum equations. Importantly, these discovered symbolic expressions not only maintain physical consistency, but also provide new interpretable insights into turbine wakes.

\subsection{Mean and instantaneous wake evolution}\label{subsec:wake}

\begin{figure*}
    \centering
    \includegraphics[width=1\linewidth]{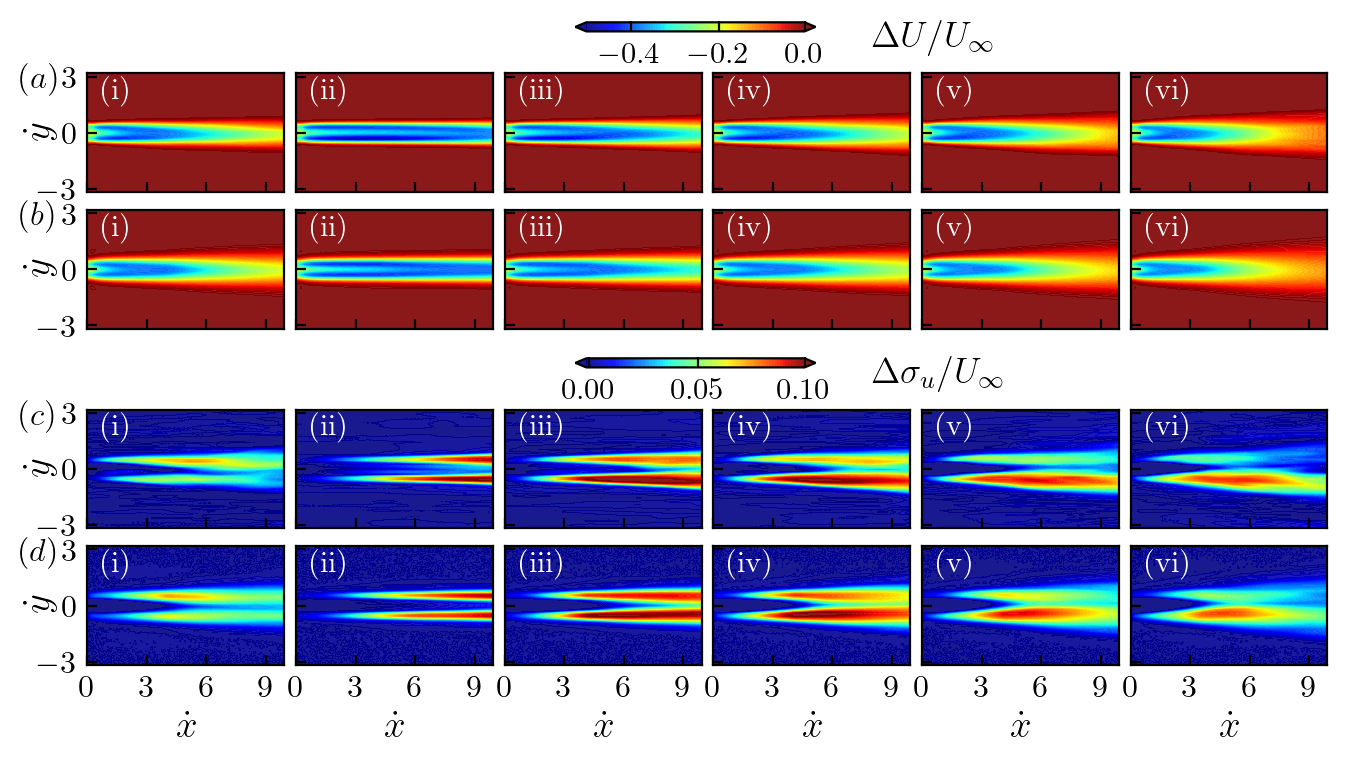}
    \caption{Comparison of (a, b) streamwise mean velocity deficit and (c, d) turbulence intensity increments from LES (a, c) and SRDWM (b, d), under varying inflow turbulence conditions: (i) PS case; (ii)-(vi) ST cases with increasing turbulence intensity.}
    \label{fig:UAndIx}
\end{figure*}

\begin{figure*}
    \centering
    \includegraphics[width=\linewidth]{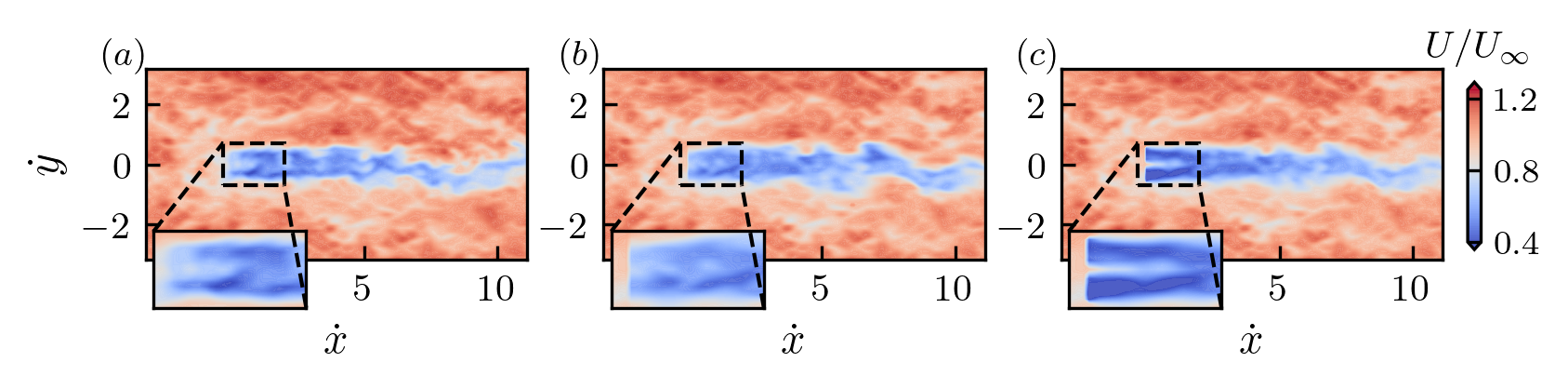}
    \caption{Instantaneous wake velocity fields from (a) LES, (b) SRDWM, and (c) conventional DWM.}
    \label{fig:instantaneous}
\end{figure*}

Figure~\ref{fig:UAndIx} presents the streamwise mean velocity deficit and turbulence intensity increments along the streamwise direction across inflow cases. These values are calculated as the differences between fields with and without the turbine. The SRDWM framework accurately reconstructs wake evolution with full spatiotemporal resolution, including both mean velocity and turbulence intensity—quantities critical to wind energy research. It shows strong robustness and generalization, automatically adapting to different inflow conditions. The quantitative comparison of errors across all cases is provided in table~\ref{tab:error}, highlighting the significant improvements of SRDWM over traditional DWM in both mean and fluctuating predictions. A notable advantage of SRDWM lies in its accurate representation of the near-wake region, which has remained a persistent challenge for conventional models. This improvement is further illustrated in figure~\ref{fig:instantaneous}, where SRDWM captures instantaneous turbulent structures more faithfully than the baseline method.

\begin{table}
\centering
\renewcommand{\arraystretch}{1.2}
\setlength{\tabcolsep}{6pt}
\begin{tabular}{llccccccccl}
\toprule
\multicolumn{2}{l}{\textbf{Case}} & PS & ST1 & ST2 & ST3 & ST4 & ST5 & Avg. & Ratio \\
\midrule
\multirow{2}{*}{$\Delta U/U_{\infty}$ ($\times 10^{-3}$)} 
& DWM (baseline) & 1.02 & 1.99 & 1.15 & 0.94 & 0.95 & 1.04 & 1.18 & \multirow{2}{*}{2.5} \\
& SRDWM (ours)   & 0.56 & 0.61 & 0.51 & 0.45 & 0.39 & 0.35 & 0.48 & \\
\midrule
\multirow{2}{*}{$\Delta \sigma_u/U_{\infty}$ ($\times 10^{-4}$)} 
& DWM (baseline) & 1.61 & 2.53 & 3.30 & 1.97 & 0.95 & 0.82 & 1.84 & \multirow{2}{*}{4.6} \\
& SRDWM (ours)   & 0.34 & 0.23 & 0.32 & 0.51 & 0.50 & 0.52 & 0.40 & \\
\bottomrule
\end{tabular}
\caption{Normalized mean squared error (MSE) of streamwise mean velocity deficit ($\Delta U/U_{\infty}$) and turbulence intensity increments ($\Delta \sigma_u/U_{\infty}$) from DWM and SRDWM, relative to LES ground truth.}
\label{tab:error}
\end{table}

Interestingly, figure~\ref{fig:UAndIx} shows a consistent asymmetry in the mean velocity deficit, with larger values on the $\dot{y} < 0$ side—attributed to turbine rotation~\citep{bastankhah2024modelling}. However, the asymmetry of turbulence intensity in the PS case differs from the ST cases, suggesting that spanwise non-uniformity in the inflow—rather than rotation alone—contributes to wake turbulence asymmetry~\citep{Chamorro2010EffectsOT}. Despite these complexities, SRDWM effectively captures wake dynamics, highlighting its adaptability. Moreover, increasing turbulence intensity accelerates wake recovery, a well established observation. Notably, ST3 shows a more pronounced turbulence intensity increase than PS, despite similar incoming turbulence intensity, indicating additional influencing factors to be explored in section~\ref{subsec:turbulence}.

\begin{figure*}
    \centering
    \includegraphics[width=1\linewidth]{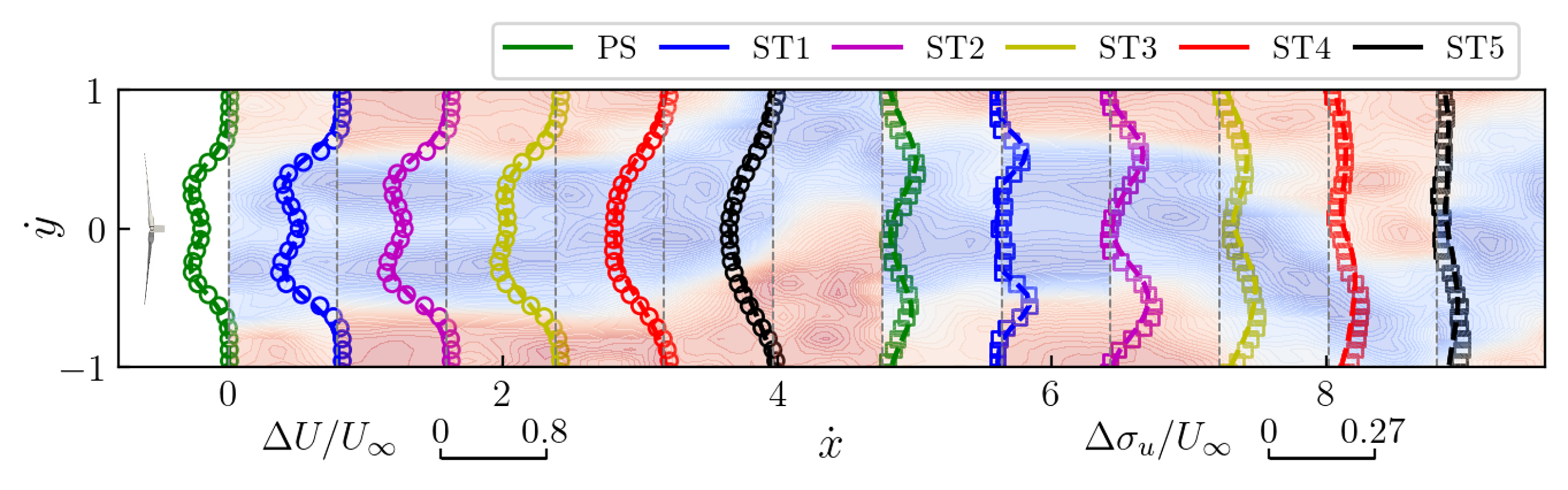}
    \caption{Profiles of streamwise mean velocity deficit (first six positions) and turbulence intensity increments (last six positions) from LES (markers) and SRDWM (lines). The twelve probe positions are equally spaced in the streamwise direction. Each profile corresponds to a fixed downstream location, comparing results for one inflow case. Background colormap shows SRDWM-computed instantaneous velocity fields for each interval.}
    \label{fig:Profiles}
\end{figure*}

Figure~\ref{fig:Profiles} offers global quantitative validation via evenly spaced downstream probes across all cases. The SRDWM predictions show good agreement with LES data in both streamwise mean velocity and wake turbulence intensity across full wake regions, including behind the rotor. The background colormaps further confirm the model’s ability to represent intensified wake meandering under higher turbulence inflow.

\begin{figure*}
    \centering
    \includegraphics[width=1\linewidth]{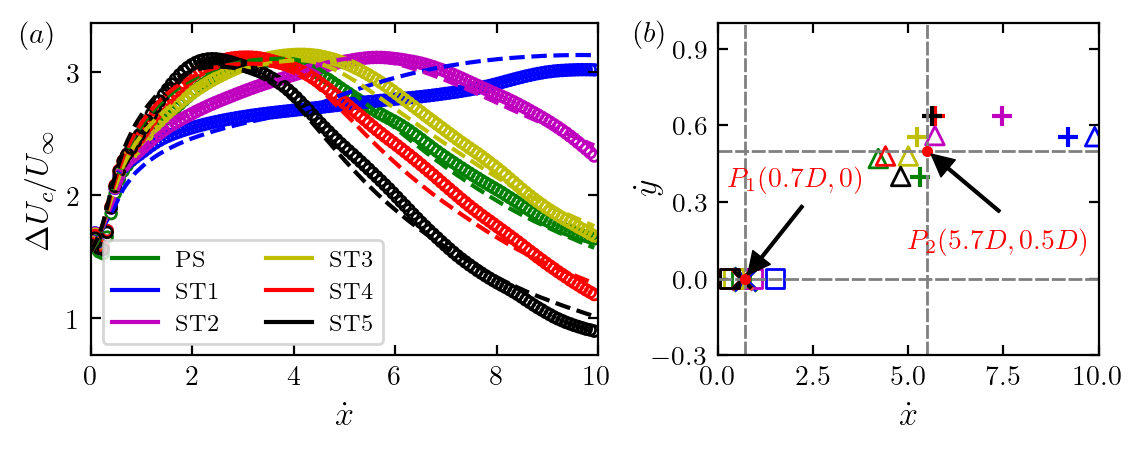}
    \caption{(a) Mean streamwise velocity deficit along the wake centerline from LES (circles) and SRDWM (lines). (b) Locations of peak rotor-extracted energy (LES: $\times$, SRDWM: $\square$) and wake-added energy (LES: $+$, SRDWM: $\triangle$). Dashed lines at $x = 0.7D$, $x = 5.7D$, $y = 0$, and $y = 0.5D$ indicate monitoring references. Red points $P_1$ and $P_2$ indicate the selected near-wake and far-wake monitoring locations for further analysis.}
    \label{fig:Uc}
\end{figure*}

To further quantify wake recovery, figure~\ref{fig:Uc}(a) presents the mean velocity deficit along the wake centerline. Enhanced inflow turbulence clearly accelerates wake recovery, whereas ST1—with very low inflow turbulence—exhibits little recovery even at $10D$ downstream. SRDWM accurately captures these trends across all conditions, with minor deviations in ST1. As shown in figure~\ref{fig:Uc}(b), turbulence intensity is suppressed in the near wake and enhanced downstream. Maximum rotor-extracted energy consistently occurs at $x = 0.7D$ along the centerline, while peak wake-added energy appears near the rotor edge ($r = 0.5D$) in the far wake. As inflow turbulence increases, this location shifts upstream, stabilizing around $x = 5.7D$. These two characteristic locations, denoted as $P_1$ and $P_2$, are therefore selected as representative monitoring points for the near- and far-wake regions in subsequent full-domain flow analysis.

\begin{figure*}
    \centering
    \includegraphics[width=1\linewidth]{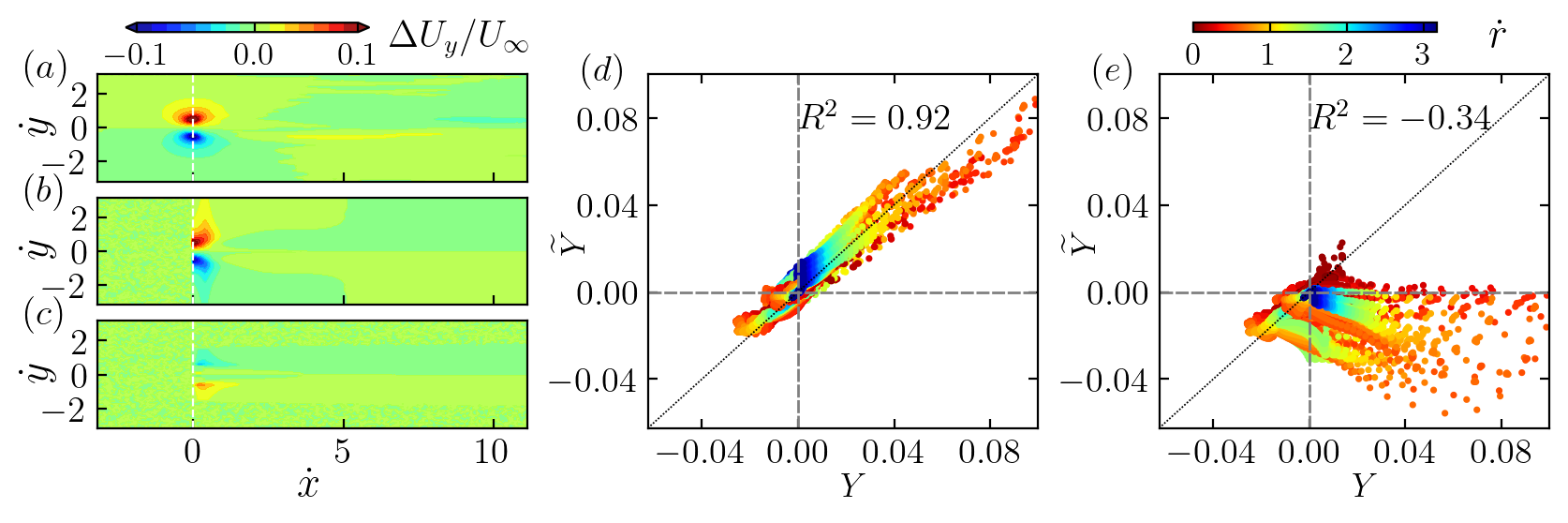}
    \caption{Mean spanwise velocity deficit in PS case from (a) LES, (b) SRDWM, and (c) conventional DWM. Identity plots comparing LES ($Y$, x-axis) and predicted ($\widetilde{Y}$, y-axis) values of normalized spanwise velocity from (d) SRDWM and (e) conventional DWM across all cases. Colors represent radial position. Dotted diagonal shows perfect match; dashed lines are zero baselines.}
    \label{fig:V}
\end{figure*}

Another major limitation of conventional DWM models lies in their incorrect representation of spanwise velocity. Figure~\ref{fig:V} compares spanwise velocity from LES, SRDWM, and the traditional DWM model. In the latter, the spanwise velocity at the boundary is fixed at zero while a large streamwise velocity deficit is imposed. As a result, solving the momentum equation yields radial velocity $U_r<0$ in the near wake, artificially compensating for this imbalance. This leads to a misrepresentation of the wake flow’s outward expansion as inward contraction, violating physical laws. In contrast, SRDWM accurately captures spanwise wake dynamics in close agreement with LES, via physically consistent boundary and forcing inputs. Identity plots in figures~\ref{fig:V}(d–e) confirm that the conventional model severely underperforms, even scoring below zero in R-squared metrics, indicating systemic directional errors. These results demonstrate the proposed framework's success in overcoming inherent deficiencies of conventional approaches.

\begin{figure*}
    \centering
    \includegraphics[width=1\linewidth]{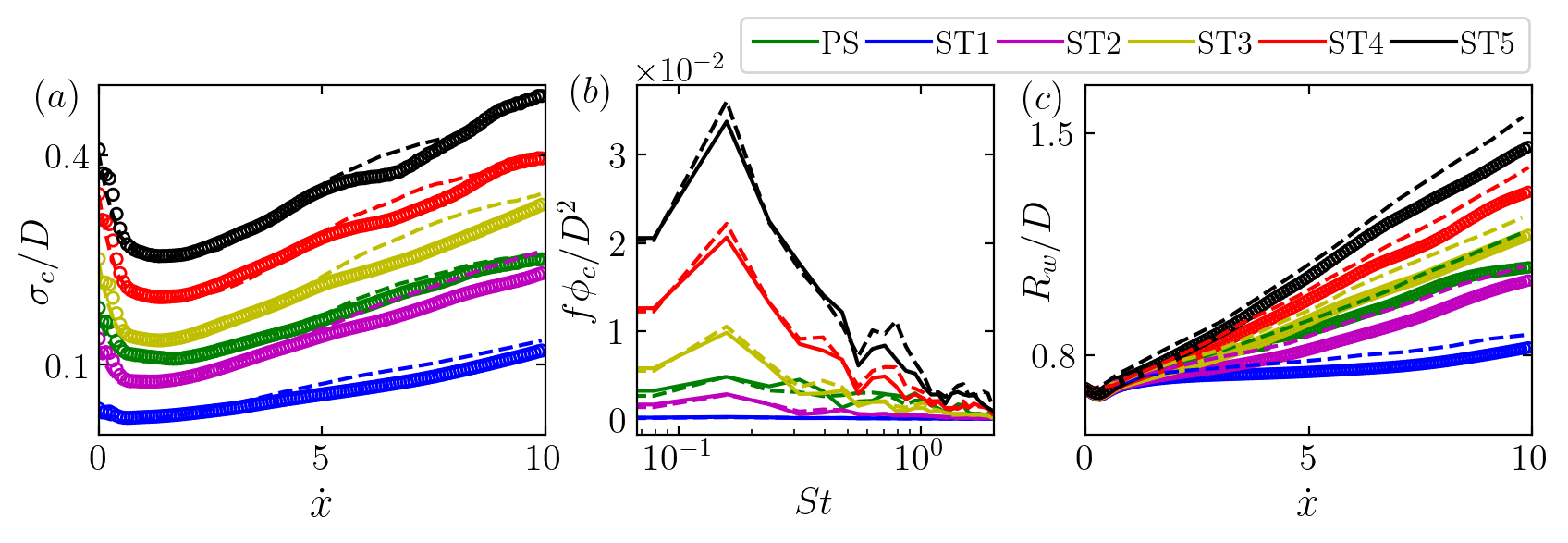}
    \caption{(a) Wake center deflection along the streamwise direction, (b) premultiplied PSD of wake center deflection at $x=0.7D$, and (c) wake width under varying turbulence intensities. SRDWM shown as dashed lines; LES as solid lines and circles.}
    \label{fig:ycAndRw}
\end{figure*}

Wake center deflection and wake width across cases are analyzed in figure~\ref{fig:ycAndRw}. Statistics of the deflection are derived by the instantaneous momentum-deficit center~\citep{Vahidi2024InfluenceOI}. A minimum wake center displacement is observed around $x=0.7D$, indicating that both minimum turbulent velocity fluctuations and wake deflection occur at this location. This finding supports the DWM hypothesis that wake meandering responds linearly to velocity fluctuations. SRDWM captures these near-wake dynamics well, further validated by the premultiplied PSD at that location (figure~\ref{fig:ycAndRw}(b)). Increased incoming turbulence intensity results in enhanced wake deflection and expansion, a trend well recovered by our model. However, wake deflection and widths are slightly overestimated, consistent with previous findings~\citep{DC_FF}, which attributed such discrepancies to intrinsic limitations of DWM-type models. Because the correlation between high-frequency fluctuations and wake meandering is generally considered negligible~\citep{Muller2015DeterminationOR}, small-scale wake deflection are filtered out in these models. Consequently, SRDWM performs less accurately for $St>0.3$, tending to overestimate wake meandering and, subsequently, the wake width.

\subsection{Turbulence statistics and wake dynamics}\label{subsec:turbulence}

\begin{figure*}
    \centering
    \includegraphics[width=1\linewidth]{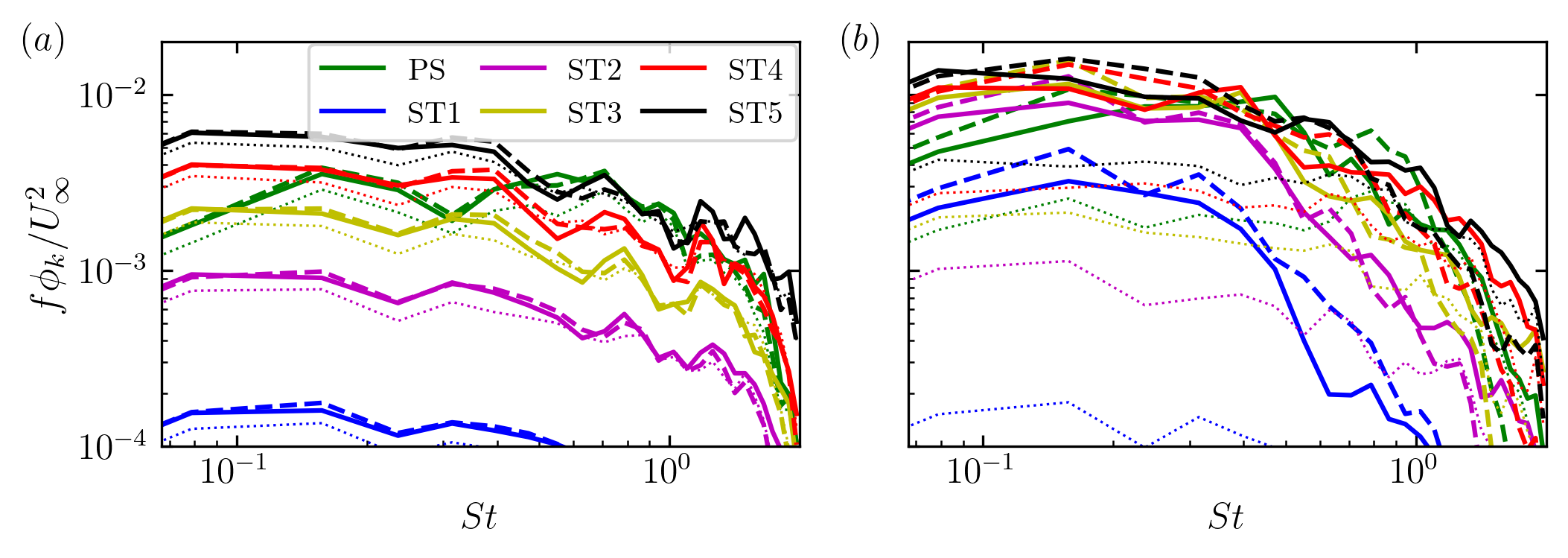}
    \caption{Premultiplied PSD of TKE along the rotor edge at (a) $x=0$ and (b) $x=5.7D$. Solid and dashed lines represent LES and SRDWM results, respectively. Dotted lines indicate cases without the turbine.}
    \label{fig:PSD}
\end{figure*}

Figure~\ref{fig:PSD} presents the premultiplied PSD of turbulence kinetic energy (TKE), defined as $f\phi_k = f (\phi_{u^{\prime}} + \phi_{v^{\prime}} + \phi_{w^{\prime}})$, where $u^{\prime}, v^{\prime}, w^{\prime}$ are velocity fluctuation components. The spectra are evaluated along the rotor edge at two downstream locations. Immediately behind the rotor ($x=0$), turbine operation enhances TKE primarily in the low-frequency range ($St < 1$), indicating the presence of large-scale coherent structures. As the flow progresses downstream ($x = 5.7D$), the influence of the incoming turbulence intensity on the low-frequency TKE gradually diminishes. In particular, lower inflow turbulence results in higher wake-added TKE contributions from the turbine, consistent with observations in figure~\ref{fig:UAndIx}. Across all cases, SRDWM reproduces turbulence energy observed in LES, confirming its ability to resolve spectral wake characteristics.

\begin{figure*}
    \centering
    \includegraphics[width=1\linewidth]{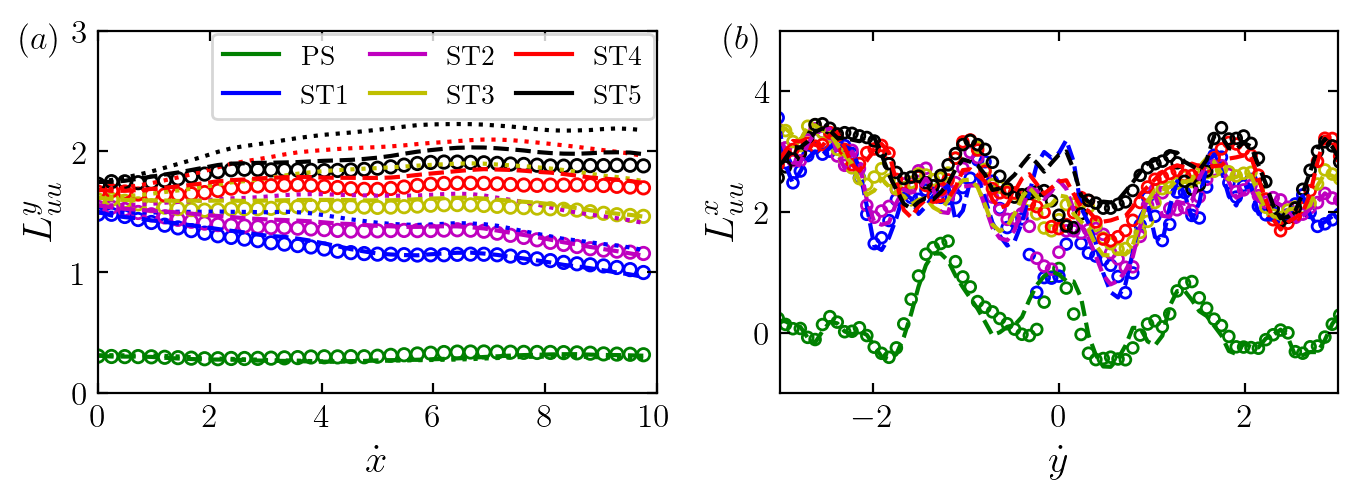}
    \caption{(a) Spanwise and (b) streamwise integral length scales of the streamwise velocity component. Circles and dashed lines represent LES and SRDWM results, respectively. Dotted lines indicate cases without the turbine.}
    \label{fig:intelength}
\end{figure*}

Notably, despite similar inflow turbulence intensities, the PS case exhibits lower wake TKE compared to the ST cases. Following the interpretation of~\cite{li2024impacts}, this phenomenon is attributed to the smaller turbulence integral length scale present in the PS condition. Their findings suggest that wake-added TKE is more pronounced when the inflow has either larger integral length scales or lower turbulence intensities. To further investigate the spatial structure of wake turbulence, we conduct an integral length scale analysis using two-point velocity correlation functions. The streamwise velocity autocorrelation is defined as~\citep{Vahidi2024InfluenceOI}:
\begin{equation}
    R_{uu}(\delta \chi) = \frac{\langle u^{\prime}(\chi) \, u^{\prime}(\chi + \delta \chi) \rangle}{\sigma_{u^{\prime}(\chi)} \, \sigma_{u^{\prime}(\chi + \delta \chi)}},
\end{equation}
where $\delta \chi$ denotes the spatial separation, and $\chi$ represents either the streamwise or spanwise direction. The corresponding integral length scale is calculated as:
\begin{equation}
    L^{\chi}_{uu} = \int_0^{\infty} R_{uu}(\delta \chi) \, d\delta \chi.
\end{equation}
As shown in figure~\ref{fig:intelength}, turbine operation reduces spatial coherence relative to no-turbine cases, reflecting large eddies breakdown into small-scale structures. SRDWM closely tracks LES across directions and inflow conditions, confirming its ability to faithfully reproduce the turbine-induced structural modifications in wakes.

\begin{figure*}
    \centering
    \includegraphics[width=1\linewidth]{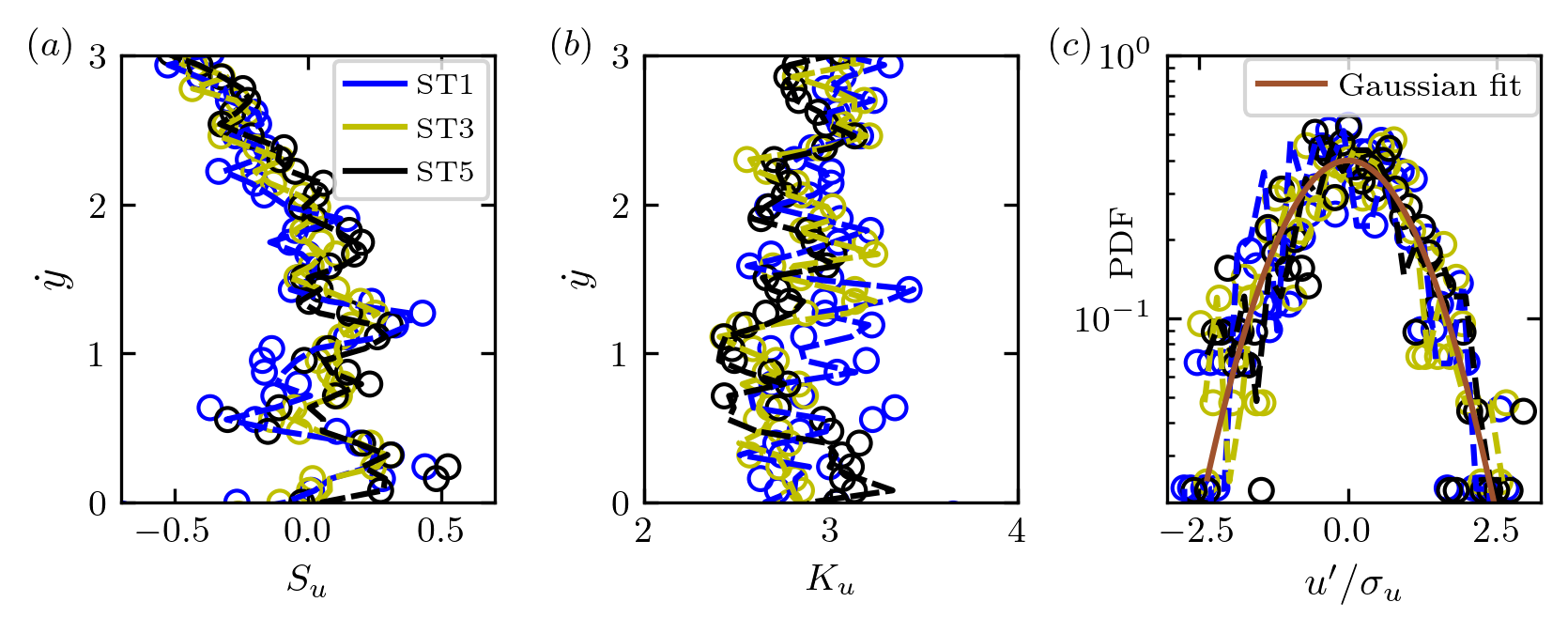}
    \caption{Spanwise profiles at $x=0.7D$ of (a) skewness, (b) kurtosis, and (c) PDF of streamwise velocity fluctuations at point $P_1$. Circles and dashed lines represent LES and SRDWM, respectively.}
    \label{fig:skewness}
\end{figure*}

High-order turbulence statistics are shown in figure~\ref{fig:skewness}. At monitoring point $P_1$, the streamwise velocity fluctuations display near-Gaussian probability distribution function (PDF), with skewness and kurtosis close to Gaussian reference values $S_u = 0$ and $K_u = 3$. These metrics are defined as:
\begin{equation}
\begin{aligned}
S_u &= \frac{\langle u^{\prime 3} \rangle}{\langle u^{\prime 2} \rangle^{3/2}}, \\
K_u &= \frac{\langle u^{\prime 4} \rangle}{\langle u^{\prime 2} \rangle^{2}}.
\end{aligned}
\label{eq:SAndK}
\end{equation}
This observation suggests that turbulence at this TKE-suppressed location is relatively homogeneous, with energy primarily distributed among continuous small-scale fluctuations rather than intermittent large-scale bursts. SRDWM accurately replicates these statistical properties, demonstrating its fidelity in capturing not only first- and second-order but also higher-order turbulence characteristics. 

\begin{figure*}
    \centering
    \includegraphics[width=1\linewidth]{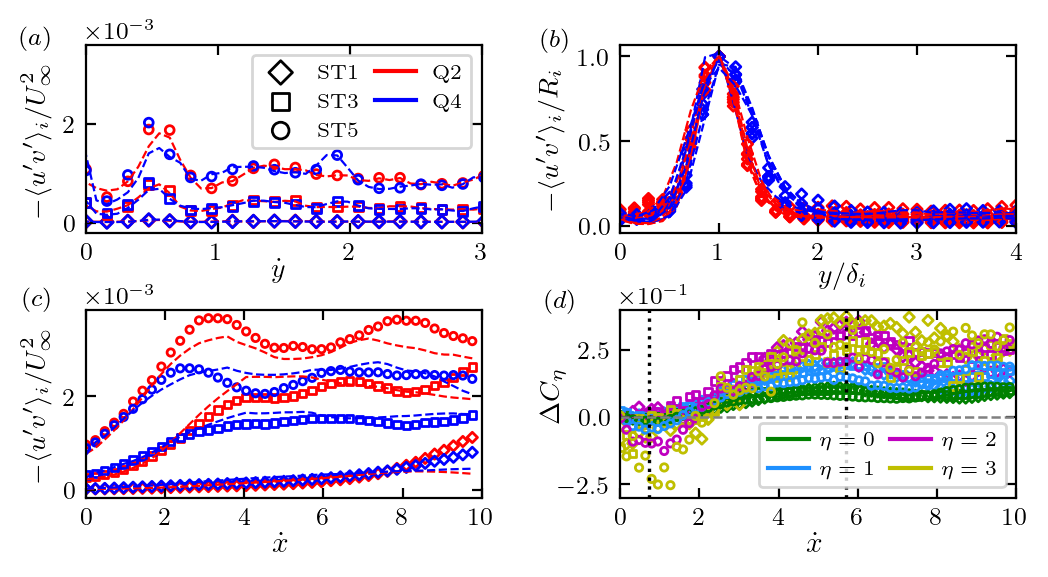}
    \caption{(a) Q2 and Q4 contributions at $x=0.7D$; (b) self-similarity of Q2/Q4 profiles for $x \geq 3D$ in ST1 case; (c) streamwise evolution of Q2 and Q4 at the rotor edge; (d) normalized difference in contribution to Reynolds stress between Q2 and Q4, $\Delta C_{\eta}=(C_{2,\eta}-C_{4,\eta})/(C_{2,\eta}+C_{4,\eta})$. Markers and dashed lines represent LES and SRDWM, respectively; dotted lines mark $x=0.7D$ and $x=5.7D$; horizontal line is zero baseline.}
    \label{fig:quadrant}
\end{figure*}

To investigate momentum transport in turbine wakes and characterize the dynamics at monitoring points $P_1$ and $P_2$, we employ quadrant analysis to quantify the contributions of coherent turbulence structures to the Reynolds shear stress~\citep{wallace2016quadrant}. The analysis is conducted in a coordinate system where the horizontal and vertical axes represent $u^{\prime}$ and $v^{\prime}$ respectively, dividing the space into four quadrants $Q_i$ that correspond to different turbulence interactions. The relative contribution of each quadrant is defined as follows:
\begin{equation}
\begin{aligned}
-\langle u^{\prime}v^{\prime} \rangle_i &= \langle -u^{\prime}v^{\prime} I_i(u^{\prime}, v^{\prime})\rangle, \\
I_i(u^{\prime}, v^{\prime}) &= 
\begin{cases}
1, & \text{if } (u^{\prime}, v^{\prime}) \text{ lies in quadrant } i, \\
0, & \text{otherwise}.
\end{cases}
\end{aligned}
\label{eq:Q}
\end{equation}
This conditional sampling method reveals the flow directionality relative to the mean flow. In figure~\ref{fig:quadrant}, we focus on the contribution of ejection events (Q2) and sweep events (Q4), which are the primary mechanisms driving wake flow transport~\citep{xiong2024self,li2024impacts}. In the near-wake region, Q2 and Q4 contribute similarly in the spanwise direction, while at the rotor edge, sweep and ejection dominate slightly inward and outward respectively. SRDWM accurately replicates these patterns, including the spanwise self-similarity of Q2 and Q4 events in low-turbulence conditions, as reported by~\cite{xiong2024self} and visualized in figure~\ref{fig:quadrant}(b). This self-similarity is observed after normalizing the peak event values and their corresponding spanwise locations at downstream positions ($x \geq 3D$). Farther downstream, wake expansion drives outdraft-related ejections, which become the dominant mechanism of momentum transport. While SRDWM effectively characterizes sweep events, it slightly underestimates the contribution of outdraft-driven ejections in the far wake, particularly under high-turbulence conditions.

To further assess the influence of intense turbulent events, we conduct quadrant-hole analysis~\citep{zhou2023large}, introducing a thresholded indicator function $I_{i,\eta}$:
\begin{equation}
    I_{i,\eta}(u^{\prime}, v^{\prime}) =
    \begin{cases}
        1, & \text{if } (u^{\prime}, v^{\prime}) \text{ lies in quadrant } i \text{ and } |u^{\prime}v^{\prime}| \geq \eta \sigma_u \sigma_v, \\
        0, & \text{otherwise}.
    \end{cases}
\end{equation}
Here, $\sigma_u$ and $\sigma_v$ denote the standard deviation of streamwise and spanwise velocity, respectively. This filtering enables isolation of high-intensity events above a prescribed threshold $\eta$. The relative contribution of such events to the total momentum flux is then expressed as:
\begin{equation}
    C_{i,\eta} = \frac{\langle u^{\prime}v^{\prime} I_{i,\eta}(u^{\prime}, v^{\prime}) \rangle}{\langle u^{\prime}v^{\prime} \rangle}.
    \label{eq:hole}
\end{equation}
As shown in figure~\ref{fig:quadrant}(d), sweep events dominate the rotor-extracted energy region, where limited outdrafts restrict momentum mixing. In contrast, wake-added energy in the far wake is primarily associated with outdrafts induced by ambient turbulence, which enhance momentum exchange between high- and low-velocity regions. This effect becomes increasingly pronounced with rising hole thresholds $\eta$, emphasizing the key role of extreme coherent structures in governing wake dynamics.

\begin{figure*}
    \centering
    \includegraphics[width=1\linewidth]{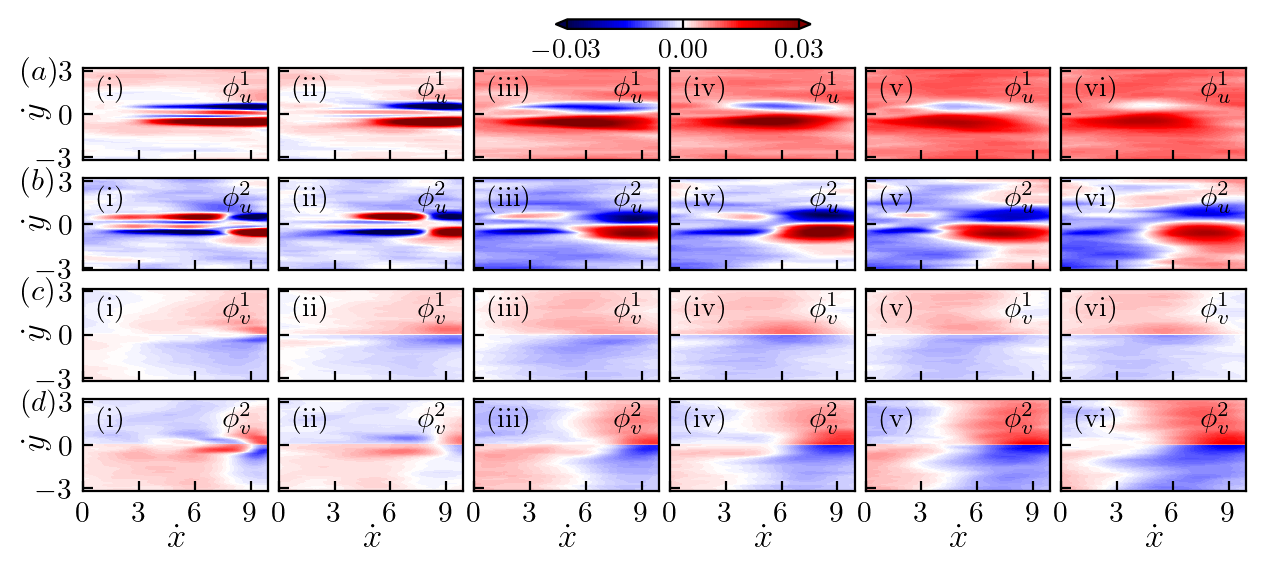}
    \caption{First two POD modes of (a, b) streamwise and (c, d) spanwise velocity. Columns (i–vi) show ST1 (i–ii), ST3 (iii–iv), and ST5 (v–vi), with LES in odd and SRDWM in even columns.}
    \label{fig:POD}
\end{figure*}

\begin{figure*}
    \centering
    \includegraphics[width=1\linewidth]{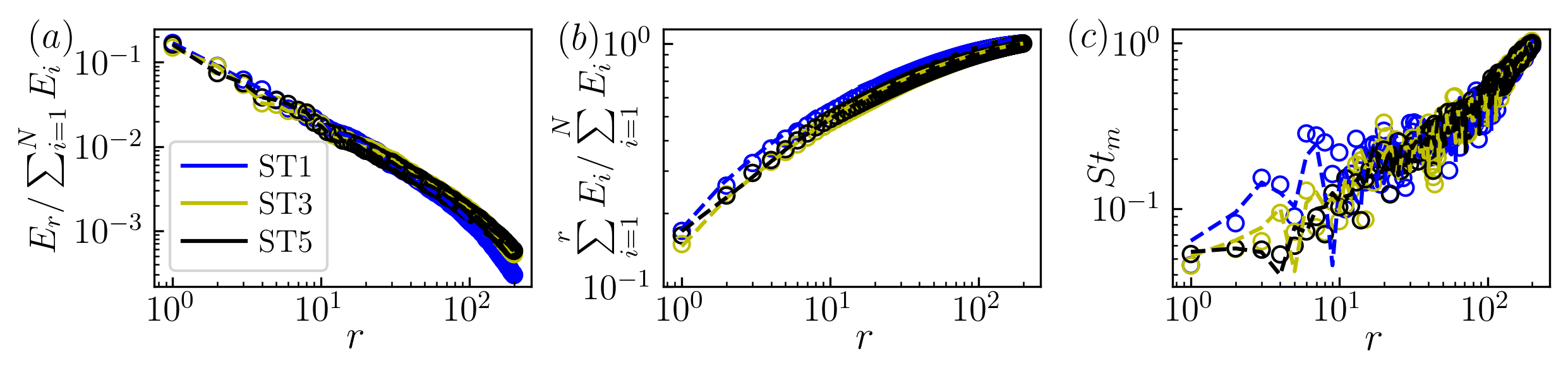}
    \caption{Variation with POD rank of (a) individual energy contribution, (b) cumulative energy, and (c) mean frequency $St_m$ of POD modes. Circles and dashed lines represent LES and SRDWM results, respectively.}
    \label{fig:POD_e}
\end{figure*}

Finally, we examine coherent turbulent structures using proper orthogonal decomposition (POD)~\citep{POD}. Figure~\ref{fig:POD} visualizes the first two POD modes for streamwise and spanwise components. SRDWM reproduces dominant large-scale wake structures observed in LES with high accuracy. The relative kinetic energy contribution of each mode is quantified in figure~\ref{fig:POD_e}, showing that modal energy distributions exhibit weak sensitivity to inflow turbulence intensity. To characterize the frequency content of individual modes, mean Strouhal number $St_m$ is defined as the geometric mean of mode frequency derived from PSD of associated temporal coefficients. Lower-order POD modes describe lower frequencies, corresponding to large-scale structures, which contain the dominant energy. Most POD modes fall within the range of $0.1 < St_m < 1$, aligning with the model’s filtering bandwidth. These results confirm that SRDWM effectively resolves the multiscale dynamics of turbine wakes.

\section{Conclusion}\label{sec:conclusion}

We have developed a Symbolic Regression-enhanced Dynamic Wake Meandering (SRDWM) model that significantly advances the capabilities of engineering wake simulations. This framework embeds symbolic expressions discovered from high-fidelity LES data, into the DWM model to reconstruct both volumetric forcing and wake boundary conditions. By incorporating these interpretable expressions, SRDWM achieves equation-level closure—enhancing physical consistency, particularly in the near wake, which has long posed a challenge for conventional approaches.

A key innovation of this work lies in the use of a hierarchical symbolic regression strategy informed by domain knowledge. By decomposing the expression space and imposing physics-informed structural constraints, we reduce the dimensionality of the learning problem and improve the interpretability and generalization of the extracted expressions. This approach supports robust equation discovery from limited data and reinforces the physical fidelity of the resulting model.

To further improve wake turbulence predictions, we introduce a revised wake-added turbulence (WAT) formulation that selectively amplifies energy-dominant fluctuations while suppressing unphysical growth near the rotor. This enables accurate reproduction of both wake spectra and turbulence statistics across a range of inflow conditions.

The SRDWM framework retains the temporal resolution capabilities of DWM while incorporating enhanced spatial integrity. It accurately captures the evolution of mean profiles and turbulent properties from the rotor to the far wake. Compared to LES, SRDWM offers comparable accuracy with runtime speedups of over three orders of magnitude. This makes it well-suited for wind farm layout optimization, control co-design, and operational forecasting. Additionally, its high computational efficiency enables the generation of large-scale wake data for training and benchmarking modern machine learning models.

These findings underscore the potential of symbolic regression as a physics-informed modeling paradigm that bridges data-driven discovery and mechanistic understanding. Unlike black-box data-driven models, SR emphasizes transparency and physical interpretability, providing deeper understanding into turbine wake dynamics. Moreover, the proposed SR framework can serve as a general tool for constructing closure models that explicitly correct empirical formulas derived under simplified assumptions, a common need across many physical domains.

Despite its advantages, the current SRDWM framework has limitations. The expressions derived from the PS case may not generalize to turbines with substantially different aerodynamic characteristics. Future work will investigate the adaptability of SR-derived forcing and boundary expressions to different turbine designs, and explore transfer learning strategies to enhance model robustness across various layouts and operating conditions.

\begin{bmhead}[Funding.]
This work was supported by National Key Research and Development Program of China (2024YFF1500600), as well as by the High Performance Computing Centers at Eastern Institute of Technology, Ningbo, and Ningbo Institute of Digital Twin.
\end{bmhead}

\begin{bmhead}[Declaration of interests.]
The authors report no conflict of interest.
\end{bmhead}

\begin{bmhead}[Author ORCIDs.]
Ding Wang, https://orcid.org/0000-0001-7006-5238; Dachuan Feng, https://orcid.org/0000-0001-5651-974X; Kangcheng Zhou, https://orcid.org/0009-0002-8874-2481; Yuntian Chen, https://orcid.org/0000-0003-4566-8197; Shijun Liao, https://orcid.org/0000-0002-2372-9502; Shiyi Chen, https://orcid.org/0000-0002-2913-4497.
\end{bmhead}




\bibliographystyle{jfm}
\bibliography{Ref-full}



\end{document}